\documentclass[pop,fleqn]{w-art}

\usepackage{times}
\usepackage{slashed,cite}
 \numberwithin{equation}{section}

\newcommand{\ft}[2]{{\textstyle\frac{#1}{#2}}}
\def\Re{\mathop{\rm Re}\nolimits}
\def\Im{\mathop{\rm Im}\nolimits}

\def\rmi{{\rm i}}

\newsavebox{\uuunit}
\sbox{\uuunit}
    {\setlength{\unitlength}{0.825em}
     \begin{picture}(0.6,0.7)
        \thinlines
        \put(0,0){\line(1,0){0.5}}
        \put(0.15,0){\line(0,1){0.7}}
        \put(0.35,0){\line(0,1){0.8}}
       \multiput(0.3,0.8)(-0.04,-0.02){12}{\rule{0.5pt}{0.5pt}}
     \end {picture}}
\newcommand {\unity}{\mathord{\!\usebox{\uuunit}}}
\newcommand{\SU}{\mathop{\rm SU}}

\newcommand{\U}{\mathop{\rm {}U}}
\newcommand{\USp}{\mathop{\rm {}USp}}

\newcommand{\Red}[1]{#1}
\newcommand{\OliveGreen}[1]{#1}

\newcommand{\RawSienna}[1]{#1}
\newcommand{\Blue}[1]{#1}

\begin{document}

\begin{titlepage}
\begin{flushright}
KUL-TF-04/30\\
hep-th/0410053
\end{flushright}
\vspace{.5cm}
\begin{center}
\baselineskip=16pt
{\LARGE   Supergravity with Fayet-Iliopoulos terms and R-symmetry 
}\\
\vfill
{\large Antoine Van Proeyen $^\dagger$
  } \\
\vfill
{\small Instituut voor Theoretische Fysica,\\ Katholieke Universiteit Leuven,\\
       Celestijnenlaan 200D\\ B-3001 Leuven, Belgium.
 }
\end{center}
\vfill
\begin{center}
{\bf Abstract}
\end{center}
{\small The simplest examples of gauged supergravities
  are $N=1$ or $N=2$ theories with Fayet-Iliopoulos (FI) terms. FI terms in supergravity
  imply that the R-symmetry is gauged. Also the $\U(1)$ or $\SU(2)$ local symmetries of K{\"a}hler and
  quaternionic-K{\"a}hler manifolds contribute to R-symmetry gauge fields. This short review clarifies the relations.
 }\vspace{2mm} \vfill \hrule width 3.cm {\footnotesize
\noindent $^\dagger$ e-mail: {\sf Antoine.VanProeyen@fys.kuleuven.ac.be}
}
\end{titlepage}
\tableofcontents{}

\section{Introduction}

In recent years, progress has been made towards building string
configurations that describe theories close to the standard model, and
models for early-universe cosmology. These involve configurations of
intersecting branes with fluxes. The effective field theory descriptions
lead to gauged supergravities. This means that the automorphism group of
the supersymmetries is gauged by a physical spin-1 gauge field of the
supergravity theory. The effective field theories for standard model
building or cosmology often involve Fayet-Iliopoulos (FI) terms in $N=1$
or $N=2$, $D=4$ supersymmetric models. In cosmological models, this
allows to raise the value of the cosmological constant and as such obtain
de Sitter vacua~\cite{Kallosh:2001tm}. FI terms are the simplest examples
of gauged supergravity theories, though the connection to gauged
R-symmetry was not always emphasized, or was even neglected by some model
builders. We recently~\cite{Binetruy:2004hh} stressed this relation. This
lead to a better understanding of some theories for inflation and for the
description of cosmic strings within supergravity~\cite{Dvali:2003zh}.
Though the latter applications were discussed during the talk at this
workshop, I will in these proceedings restrict myself to the explanation
of R-symmetry and FI terms. I will include also $N=2$ theories in this
text, as the occurrence of R-symmetry for $N=2$ is very similar to the
situation for $N=1$.

The automorphisms of the supersymmetries in various dimensions are
determined by properties of the Clifford algebra. In 4 dimensions with
$N$ supersymmetries the group is $\U(N)$, i.e. it is $\U(1)$ for $N=1$
and $\SU(2)\times \U(1)$ for $N=2$. These $N=2$ theories exist in very
similar forms in 5 and 6 dimensions where the automorphism group is
$\USp(2)=\SU(2)$. In 4 dimensions, the supersymmetry parameters can be
split in 2 chiral parts $\epsilon =\epsilon _L+\epsilon _R$, which are
essentially each others complex conjugates (strictly speaking: charge
conjugates). The $\U(1)$ R-symmetry is a phase transformation
$\epsilon\rightarrow \exp\left(\rmi \alpha\gamma _5\right)\epsilon $ or
$\epsilon _L\rightarrow \exp\left(\rmi \alpha \right)\epsilon _L$.

In section \ref{ss:N1SG}, a recapitulation of the basic ingredients of
matter-coupled $N=1$ supergravity is given~\cite{Cremmer:1983en}. It is
indicated which data determine the theory completely, and where the FI
term fits in this description. A short summary is also given for $N=2$
supergravity. The superconformal formulation clarifies the structure of
many equations, and in particular of the connection of R-symmetry with
the FI term and other gauged symmetries. This is explained in section
\ref{ss:SC}. Finally, in section \ref{ss:geometricR} the relation of
R-symmetry with symmetries in the geometry of the scalar manifolds is
clarified. This involves the connections of K{\"a}hler and
quaternionic-K{\"a}hler manifolds.

\section{Matter-coupled $N=1$ and $N=2$ supergravity}
\label{ss:N1SG}

Pure $N=1$ supergravity contains a spin 2 and a spin 3/2 field: the
graviton and the gravitino. One can add a number of vector multiplets,
each having a spin 1 and a spin 1/2 field. The spin 1 fields that are
introduced in this way are gauge fields for an arbitrary ordinary gauge
group. Chiral multiplets contain a complex spin 0 field and a spin 1/2
field. These chiral multiplets can transform in a representation of the
gauge group. I will restrict myself here to these multiplets, though
recently it was emphasized \cite{Grimm:2004uq} that tensor (linear)
multiplets are a natural description for some string compactifications. A
description of the couplings of tensor multiplets is given
in~\cite{Binetruy:2000zx}. It was shown already in~\cite{Ferrara:1983dh}
that the tensor multiplet couplings in $N=1$ supergravity can be dualized
to couplings of chiral multiplets. It is thus sufficient to restrict to
chiral, vector and the pure supergravity multiplets to have a complete
description of $N=1$ matter-coupled supergravities, though of course the
tensor multiplet description may be more useful for some applications.
Let me still emphasize that the discussion is restricted here to actions
with at most terms with two spacetime derivatives.

The supersymmetry transformations of the fermions of supergravity
theories contain a lot of information. The graviton is described by the
vierbein $e_\mu ^a$, and its fermionic partner is the gravitino $\psi
_\mu $. The bosonic part of the supersymmetry transformation of the
latter is
\begin{equation}
  \Blue{\delta} \Red{\psi _{\mu L}}   =  \left[ \partial _\mu  +\ft14
\OliveGreen{\omega _\mu {}^{ab}(e)}\gamma _{ab} +\ft 12\rmi
\OliveGreen{A_\mu^B} \right]\Blue{\epsilon_L} +\ft12 \kappa^2\gamma _\mu
\OliveGreen{F_0} \Blue{\epsilon _R},
 \label{susypsimu}
\end{equation}
where $M_P=\kappa ^{-1}$ is the Planck mass. There are 3 objects that
appear in this transformation law. The first one is the spin-connection
$\omega _\mu {}^{ab}$, which depends on the graviton $e_\mu ^a$. The
second one is $A_\mu^B$, a composite gauge field for the $\U(1)$
R-symmetry as will be explained below. The third one is an auxiliary
scalar $F_0$. Both $A_\mu^B$ and $F_0$ depend on fields of other
multiplets.

The vector multiplets contain the gauge vector fields $W_\mu ^\alpha $,
with gauge field strengths $F_{\mu \nu}^\alpha$, and the `gaugini'
$\lambda ^\alpha $.
The chiral multiplets have complex scalars $\phi _i$ (we denote the
complex conjugates as $\phi ^i$) and the matter fermions, whose
left-chiral parts are denoted as $\chi _i$. The fermions transform under
supersymmetry as (again only bosonic terms)
\begin{equation}
  \Blue{\delta} \Red{\lambda^\alpha} =\ft14\gamma ^{\mu \nu }
  \OliveGreen{F_{\mu \nu}^\alpha } \Blue{\epsilon}
  +\ft12\rmi \gamma _5 \OliveGreen{D}^\alpha  \Blue{\epsilon},\qquad
\Blue{\delta} \Red{\chi _i} =  \ft12\hat{\slashed{\partial}
}\OliveGreen{\phi_i}\Blue{\epsilon _R}- \ft12
\OliveGreen{F_i}\Blue{\epsilon _L}. \label{susyfermions}
\end{equation}
Here appear auxiliary scalars $D^\alpha $ and $F_i$, which are functions
of the physical fields.

Considering the bosonic part of the action
\begin{eqnarray}
e^{-1}{\cal L}_{\rm bos}&=&-\ft12\kappa ^{-2} R -\ft14 (\Re
\Blue{f_{\alpha \beta}})F_{\mu \nu }^\alpha F^{\mu \nu \,\beta } +\ft
14\rmi(\Im \Blue{f_{\alpha \beta}})
 F_{\mu \nu }^\alpha \tilde F^{\mu \nu \,\beta }\nonumber\\
&& -\Blue{g_i{}^j}(\hat{\partial }_\mu \phi ^i)(\hat{\partial }^\mu \phi
_j) -V(\phi ,\phi ^*),
\end{eqnarray}
one can enumerate the objects that completely determine the theory:
\vspace{-2mm}
\begin{enumerate}
  \item the gauge group with generators enumerated by the values of the
  index $\alpha $. The kinetic energy of the gauge fields is determined
  by holomorphic functions $f_{\alpha \beta }(\phi )$.
 \vspace{-2mm}
  \item a representation of this gauge group enumerated by the values of
  the index $i$. The kinetic energy of the scalars is determined by a
  real K{\"a}hler potential $K(\phi ,\phi ^*)$, up to an equivalence:
\begin{equation}
  \Blue{g_i{}^j=}g^j{}_i=\partial _i\partial ^j \Blue{K}(\phi ,\phi ^*), \qquad
\Blue{K}(\phi ,\phi ^*)\approx\Blue{K'}(\phi ,\phi ^*)=\Blue{K}(\phi
,\phi ^*)+ f(\phi)+f^*(\phi^*).
 \label{gfromKahler}
\end{equation}
\vspace{-6mm}
  \item a holomorphic superpotential $W(\phi )$, which will determine
  the potential $V(\phi ,\phi ^*)$, see below.
  \vspace{-2mm}
  \item for any $\U(1)$ factor in the gauge group: a constant $\xi _\alpha
  $ (the `FI constant').
\end{enumerate}
It is a general result of supergravity that the potential is given by the
square of the supersymmetry transformations of the fermions, where the
kinetic matrix determines how this square is formed. For the case of
$N=1$, this leads to
\begin{equation}
  V= -3\kappa^2\OliveGreen{F_0F^0}+\OliveGreen{F_i}g^i{}_j\OliveGreen{F^j} +\ft12
\OliveGreen{D^\alpha}(\Re f_{\alpha \beta}) \OliveGreen{D^\beta},\qquad
  \OliveGreen{D^\alpha}   = (\Re \Blue{f})^{-1\,\alpha \beta }  \OliveGreen{{\cal
  P}_\beta}.
 \label{VN1}
\end{equation}
One recognizes here the auxiliary scalar expressions that appear in the
fermion transformations (\ref{susypsimu}), (\ref{susyfermions}). The
first two terms together are called the `$F$-term', while the last one is
called the `$D$-term'. The $F$-term depends on the superpotential $W$,
and on the K{\"a}hler potential $K$ (order $\kappa^2$ corrections). We do not
need its explicit expression here. The $D$-term, on the other hand,
depends on the gauge transformations and on $K$. Moreover, this contains
the FI term, depending on the above-mentioned FI constants. The quantity
${\cal P}_\alpha $ is the `moment map' of the symmetry indicated by the
index $\alpha $. It has a geometric meaning that we will discuss further
in section \ref{ss:geometricR}. The value of this moment map involves the
FI constants. However, the addition of FI term gives restrictions in
supergravity  (e.g. on the superpotential). These have been neglected in
previous papers on D-term cosmology~\cite{Binetruy:1996xj}. One of the
aims of~\cite{Binetruy:2004hh} is to put this straight! The restriction
on the superpotential will be explained in section \ref{ss:SC}.

That the $D$-term is related to R-symmetry is already clear by the fact
that the moment map appears also in the expression of the composite
$\U(1)$ gauge field that we encountered in (\ref{susypsimu}):
\begin{equation}
  A_\mu ^B=\omega ^i\partial _\mu \phi_i+\omega _i\partial _\mu \phi^i
  +\kappa^2W_\mu ^\alpha \OliveGreen{{\cal P}_\alpha}, \qquad
  \omega ^i\equiv -\ft12\rmi \kappa^2 \partial ^iK,\qquad
  \omega _i\equiv  \ft12\rmi \kappa^2 \partial _iK.
 \label{valueAmuB}
\end{equation}
Denoting the gauge transformations of the scalars as
\begin{equation}
  \delta _G\phi_i=\Lambda ^\alpha k_{\alpha i}(\phi),\qquad\delta _G\phi^i=\Lambda ^\alpha
  k_\alpha^i(\phi^*),
 \label{delGz}
\end{equation}
the value of the moment map can be written as
\begin{equation}
   \OliveGreen{{\cal P}_\alpha }
   = \ft12\rmi k_{\alpha i}\partial ^iK-\ft12\rmi k_\alpha^i\partial_iK
   - \ft32\rmi \tilde r_\alpha (\phi )
   + \ft32\rmi \tilde r_\alpha ^*(\phi^*).
 \label{valueP}
\end{equation}
These gauge symmetries do not necessarily leave the K{\"a}hler potential
invariant, but due to the equivalence shown in (\ref{gfromKahler}), it
can transform as the real part of a holomorphic function. That determines
the value of the quantities $r_\alpha (\phi )$ in (\ref{valueP}):
\begin{equation}
   \delta _G \Blue{K}\equiv \Lambda ^\alpha \left( k_{\alpha i} \partial ^i\Blue{K}+
 k_\alpha^i \partial _i\Blue{K}\right)
 = 3\Lambda ^\alpha \left([\tilde r_\alpha (\phi )
                  +\tilde r^*_\alpha(\phi ^*)\right].
 \label{delalphaK}
\end{equation}
However, the above formula does not determine $r_\alpha (\phi )$
completely as only its real part occurs. As these functions should be
holomorphic, the only remaining freedom are imaginary constants:
\begin{equation}
\tilde r_\alpha = \ldots + \ft13\rmi  \RawSienna{\xi_\alpha }.
 \label{ralphaFI}
\end{equation}
To be consistent with the gauge group, this addition is only possible for
Abelian factors, as can be understood from the relations in the next
section, see (\ref{equivN1}).

Observe the 3 different types of terms contributing to the R-symmetry
gauge field. First, there are the field-dependent shifts of bosons. The
first terms of (\ref{valueAmuB}) are the pull-back of the K{\"a}hler
connection, which will be further explained in section
\ref{ss:geometricR}. The first terms in (\ref{valueP}) are the
covariantizations of these terms. Secondly, there are the possible
contributions from a non-invariant K{\"a}hler potential, and finally there is
the possibility of the constant FI terms. In string theory there are no
free constants. E.g. coupling constants appear as vacuum expectation
values of moduli fields. A possible way in which FI terms may appear is
that one considers an effective theory in which some chiral multiplets
are already integrated out. In the full theory with these chiral
multiplets, the moment map may be non-zero and field-dependent. After
fixing the fields that are integrated out, the resulting value of the
moment map may be a constant that remains in the effective theory as a FI
constant. This has been illustrated in more detail
in~\cite{Binetruy:2004hh}.

Finally, let us consider the analogous facts for $N=2$. The facts that we
mention here are as well applicable to
$D=4$~\cite{deWit:1984pk,deWit:1985px,Andrianopoli:1997cm},
$D=5$~\cite{Ceresole:2000jd,Bergshoeff:2004kh} or
$D=6$~\cite{Bergshoeff:1986mz}. We consider theories with $n_V$ vector
multiplets and $n_H$ hypermultiplets. The former contain gauge fields
$A_\mu ^I$ for a gauge group, and there is one extra gauge field, the
`graviphoton', part of the pure supergravity multiplet, such that
$I=0,1,\ldots ,n_V$. In 4 and 5 dimensions there are further complex or
real scalar fields, which are not important for the R-symmetry story. The
hypermultiplets contain each 4 real scalars that combine in quaternions.
The scalar manifold is a quaternionic-K{\"a}hler manifold (some aspects of
these manifolds are reviewed in section \ref{ss:geometricR}). We use here
a basis of real scalars $q^X$, with $X=1,\ldots ,4n_H$. These may
transform in a representation of the gauge group defined by the vector
multiplets. The supersymmetry transformation of the gravitini contains a
composite gauge field for the $\SU(2)$ group in the automorphism group of
the supersymmetries. Its value is very similar to (\ref{valueAmuB}):
\begin{equation}
  \vec{{\cal V}}_\mu =\partial _\mu q^X\vec{\omega }_X +\kappa^2A_\mu
  ^I\vec{P}_I,
 \label{auxVmu}
\end{equation}
where $\vec{\omega }_X$ is a connection on the quaternionic-K{\"a}hler
manifold determined by the scalars of the hypermultiplets. On the other
hand, we have here a triplet moment map $\vec{P}_I$, related to the fact
that there are 3 complex structures $\vec{J}_X{}^Y$ on the
quaternionic-K{\"a}hler manifold, while there is only 1 on the K{\"a}hler
manifold, see section \ref{ss:geometricR}. Using the notation
 $\delta q^X= -\Lambda ^Ik_I^X$ (the minus
sign is chosen for consistency with
\cite{Ceresole:2000jd,Bergshoeff:2004kh}), one finds for these moment
maps:
\begin{equation}
  4n_H\kappa^2 \vec{P}_I=\vec{J}_X{}^Y D_Yk_I^X.
 \label{valuePN2}
\end{equation}
The value is undetermined when there are no hypermultiplets ($n_H=0$) or
in rigid supersymmetry ($\kappa=0$). In these cases $\vec{P}_I$ can be
constants and the gauge symmetry implies that the corresponding gauge
group should be $\U(1)$ or $\SU(2)$.

\section{The superconformal origin of R-symmetry}
\label{ss:SC}

The structure of matter-coupled supergravities is clarified by using a
superconformal approach. This was developed for $N=1$
in~\cite{Ferrara:1977ij,Kaku:1978ea}, and a recent convenient summary is
in~\cite{Kallosh:2000ve}. For $N=2$
see~\cite{deWit:1985px,Bergshoeff:1986mz,Bergshoeff:2004kh}. One first
considers actions invariant under the superconformal group. This
contains, apart from the super-Poincar{\'e} group, also dilatations and
special conformal transformations, special supersymmetry and a local
R-symmetry group. After gauge fixing of all these extra symmetries, the
remaining theory is an ordinary super-Poincar{\'e} theory. Concerning the
fields, one starts from the so-called Weyl multiplet. This contains gauge
fields for all the superconformal symmetries. Some of these gauge fields
will not be independent fields, but functions of the physical fields.
E.g. in this multiplet will be a gauge field for the $\U(1)$
transformation that acts on the gravitino,
 $
  \delta _{\U(1)}\psi _\mu =-\ft12\rmi\gamma _5\Lambda _{\U(1)}\psi _\mu$.

Apart from the physical multiplets, mentioned in the previous section,
there is also an extra chiral multiplet that contains non-physical fields
whose values are gauge-fixed in order to break the superconformal group
at the end to the super-Poincar{\'e} group. The corresponding scalar is
called the `conformon' $Y$. Fixing its value introduces the Planck mass
in the theory that was first of all conformal invariant. The gauge
condition for dilatations fixes the modulus, while the phase of $Y$ is
fixed by a gauge choice for the $\U(1)$ R-symmetry in the superconformal
group:
\begin{equation}
  \mbox{Dilatation gauge:}\quad  YY^* {\rm e}^{-\kappa ^2 K/3}= \kappa ^{-2},\qquad
  \U(1)\mbox{-gauge:}\quad Y=Y^*.
 \label{DU1gauge}
\end{equation}
In general, one thus starts with $n+1$ chiral multiplets, whose rigid
symmetries define a rigid K{\"a}hler manifold. The couplings in this larger
scalar manifold are restricted by the presence of a conformal symmetry.
After gauge fixing, the remaining K{\"a}hler manifold has complex dimension
$n$. The notation for the auxiliary field $F_0$ and the inclusion of the
first term in (\ref{VN1}) in the `F-term' potential finds here its place,
as they are related to the transformation of the fermion of the
compensating chiral multiplet.

Without vector multiplets, there are still the first two terms in
(\ref{valueAmuB}). But there is no gauge invariance left. The $\U(1)$ of
the superconformal group is broken by the gauge choice (\ref{DU1gauge}).
This is called `ungauged supergravity'. `Gauged supergravity' is the
situation in which physical fields enter in the expression $A_\mu ^B$.
Hence, it occurs when ${\cal P}_\alpha \neq 0$. This is the case when
chiral multiplets transform non-trivially under the gauge group or when
FI terms are added.

Therefore, we now consider the gauging of symmetries in the
superconformal formulation. Apart from the scalars $\phi_i$, whose
transformations were given in (\ref{delGz}), there is now also the
conformon. Its transformation law is parametrized as
\begin{equation}
  \delta _G Y= Y r_\alpha (\phi) \Lambda ^\alpha,\qquad r_\alpha =\kappa^2\tilde
  r_\alpha.
 \label{delalphaY}
\end{equation}
The way in which $Y$ appears in the right-hand side is imposed by the
fact that the gauge group should commute with the dilatations. The
notation $r_\alpha $ is chosen, because one can check that with the
dilatation gauge condition (\ref{DU1gauge}), this transformation of $Y$
corresponds to the transformation (\ref{delalphaK}) for the K{\"a}hler
potential. From (\ref{ralphaFI}) one sees that the FI term appear when
the conformon undergoes a phase transformation under the gauge symmetry.

The $\U(1)$ gauge choice (\ref{DU1gauge}) is not invariant under gauge
transformations. This implies that in the final theory the gauge
transformations get an extra contribution from the superconformal
$\U(1)$. The preservation of this gauge condition gives
\begin{equation}
  \Lambda _{\U(1)}=\ft32\Lambda ^\alpha(r^*_\alpha -r_\alpha )=\Lambda ^\alpha
  \left( \kappa^2{\cal P}_\alpha +\omega ^ik_{\alpha i}+\omega _i
  k_\alpha^i\right).
 \label{U1togauge}
\end{equation}
This implies that the gravitino now transforms under all the gauge
transformations for which $r_\alpha \neq 0$. One can compute the gauge
transformations of the auxiliary field (\ref{valueAmuB}) and finds that
the gauge transformations of the fields $\phi_i$ and $W_\mu ^\alpha $
induce
\begin{equation}
  \delta _GA_\mu ^B=
  \partial _\mu\Lambda _{\U(1)}
 \label{delGAmuB}
\end{equation}
To prove this equation one has to use the `equivariance condition' that
follows from the fact that (\ref{delalphaY}) satisfies the gauge algebra:
\begin{equation}
   k_\alpha^jg_j{}^ik_{\beta i} - k_\beta^jg_j{}^ik_{\alpha i}
+\rmi f_{\alpha \beta }{}^\gamma {\cal P}_\gamma =0.
 \label{equivN1}
\end{equation}

The dilatational invariance imposes that the superpotential in the
superconformal theory should be of the form
\begin{equation}
  {\cal W}=Y^3 \kappa^3 W(\phi).
 \label{conformalW}
\end{equation}
This superpotential should be invariant under the gauge symmetries. In
view of (\ref{delalphaY}) this implies that $W(\phi)$ should transform
homogeneously under the gauge group
\begin{equation}
  \delta _G W\equiv (\partial ^iW) \Lambda ^\alpha k_{\alpha i}=-3\Lambda ^\alpha
  r_\alpha (\phi)W(\phi).
 \label{delalphaW}
\end{equation}
This is important for the cosmological scenarios that go under the name
of $D$-term inflation~\cite{Binetruy:1996xj,Halyo:1996pp}. Also the KKLT
scenario~\cite{Kachru:2003aw} has $D$-terms in its effective supergravity
description. E.g. (\ref{delalphaW}) implies that when one introduces FI
terms for a symmetry, the superpotential should have a non-vanishing
homogeneous transformation under this symmetry.

In $N=2$ supergravity one has two extra multiplets in the superconformal
approach. First there is an extra vector multiplet that contains the
graviphoton apart from the conformon and its fermionic partners.
Moreover, there is a compensating hypermultiplet. To describe $n_H$
physical hypermultiplets, one starts with a hyper-K{\"a}hler manifold of
quaternionic dimension $n_H+1$ with conformal symmetry. This leads to a
super-Poincar{\'e} theory with a quaternionic-K{\"a}hler manifold of dimension
$n_H$. The quaternionic and conformal structure in this case completely
determines the moment maps. In the case $n_H=0$, the compensating
hypermultiplet may still transform under the $\U(1)$ or $\SU(2)$ gauge
symmetry. This is the origin of the FI terms similar to the phase
transformation of the conformon for $N=1$. A difference, however, is that
$N=1$ FI terms are still possible with non-trivial chiral multiplets,
while for $N=2$ supergravity the conformal and quaternionic structures do
not allow arbitrary FI terms when there are non-trivial hypermultiplets.
Note that in rigid $N=2$ supersymmetry a FI term may still be added,
because the restriction is intimately related to the presence of the
conformal symmetry that one needs to get a supergravity theory out of the
couplings of the hyper-K{\"a}hler manifold. The $\SU(2)$ part of the
superconformal group is gauge-fixed by fixing phases of the quaternion of
the compensating multiplet. Any tri-holomorphic isometry on the
$(n_H+1)$-dimensional hyper-K{\"a}hler manifold will also act on these
phases. The gauge fixing then implies again that in the Poincar{\'e}
supergravity theory, the gauge symmetries get contributions from the
$\SU(2)$ R-symmetry~\cite{Bergshoeff:2004kh} similar to
(\ref{U1togauge}):
\begin{equation}
\vec \Lambda_{SU(2)} =-g \vec \omega_X k_I^X \Lambda ^I+ g \Lambda^I \vec
P_I. \label{LambdaSU2}
\end{equation}
The auxiliary gauge field for the $\SU(2)$ symmetry that appears in the
supersymmetry transformation of the gaugini, (\ref{auxVmu}) transforms
under the gauge symmetries as an $\SU(2)$ gauge vector with parameter
(\ref{LambdaSU2}), again due to the `equivariance condition'
\begin{equation}
  2\kappa^2 \vec P_I \times \vec P_J + \ft12k_I^X\vec {J} _X{}^Yg_{YZ}
   k_J^Z- f_{IJ}{}^K \vec P_K=0.
 \label{equivariance} 
\end{equation}

\section{K{\"a}hler and quaternionic-K{\"a}hler connections related to R-symmetry}
\label{ss:geometricR}

The R-symmetry is related to reparametrizations of geometric properties
of the scalar manifold. In this final section, I give a recapitulation of
the relevant facts. The supersymmetry algebra imposes that these
manifolds have complex structures. For the chiral multiplets I adopted
already a notation in which all the coordinates, say $\phi ^x$, are split
in holomorphic coordinates $\phi ^x=\left\{ \phi_i,\phi^i\right\} $. This
means that the complex structure is diagonalized:
\begin{equation}
  J_x{}^y= \begin{pmatrix}J^i{}_j&0\\ 0&J_i{}^j\end{pmatrix},\qquad
  J^i{}_j=-J_i{}^j=-\rmi \delta_i^j,\qquad J_x{}^yJ_y{}^z=-\delta _x^z.
 \label{N1complexstructure}
\end{equation}
The scalars of the hypermultiplets in $N=2$, denoted by $q^X$, form a
manifold with a quaternionic structure, i.e. there is the triplet
$\vec{J}_X{}^Y$ such that for arbitrary 3-vectors $\vec \alpha $ and
$\vec \beta $,
\begin{equation} \label{quatalg}
\vec \alpha \cdot \vec J\,\vec \beta \cdot \vec  J = -\unity _{4r} \vec
\alpha \cdot \vec \beta + \left( \vec \alpha \times \vec \beta\right)
\cdot \vec J.
\end{equation}
Another Ward identity of supergravity implies that these complex
structures are related to the curvature of the connection that appears in
the gravitino transformation law, see e.g. for $N=1$:  (\ref{susypsimu}),
(\ref{valueAmuB}). The expressions are:
\begin{eqnarray}
  N=1& :& R_i{}^j\equiv \partial _i\omega ^j-\partial ^j\omega _i=
  -\kappa^2J_i{}^kg_k{}^j,\qquad
R^i{}_j\equiv \partial^i\omega_j-\partial _j\omega ^i=
  -\kappa^2J^i{}_kg^k{}_j,
  \nonumber\\
  N=2& : & \vec{R}_{XY}\equiv  2\partial_{[X}\vec \omega _{Y]}
+2\vec \omega _X\times \vec \omega _Y=-\ft12\kappa^2\vec{J}_X{}^Zg_{ZY}.
 \label{curvJ}
\end{eqnarray}
Gauged isometries have to preserve the complex structures. They are
therefore holomorphic in $N=1$, and for the hypermultiplets of $N=2$
there is a similar requirement that they preserve the quaternionic
structure. In each case it can be shown that this implies the existence
of moment maps:
\begin{equation}
 N=1 \ : \ \partial _i{\cal P}_\alpha =J_i{}^jg_j{}^kk_{\alpha k}, \qquad
 N=2 \ : \     D_X \vec{P}_I\equiv \partial
  _X\vec{P}_I+2\vec{\omega }_X\times\vec{P}_I = -\ft12\vec{J}_{XY}k_I^Y.
 \label{defMomentMaps}
\end{equation}
These moment maps gives the amount by which a gauge symmetry contributes
to the R-symmetry. We saw that the value of the auxiliary $D$-field is
proportional to it, see (\ref{VN1}) and also in $N=2$ a triplet auxiliary
field of vector multiplets is proportional to $\vec{P}_I$. For $N=1$ this
moment map is only determined up to a constant, and the latter is then
the FI-term. For $N=2$ the curvature of the quaternionic-K{\"a}hler manifold
does not allow this, and FI terms are only possible without physical
hypermultiplets in supergravity.

For $N=1$ one uses the K{\"a}hler potential to describe the manifold. The
latter is not uniquely defined, but can transform under `K{\"a}hler
transformations' as in (\ref{gfromKahler}). In the conformal approach
these originate from a change of definition of the conformon from $Y$ to
$Y'=Y\exp[\kappa ^2f(\phi)/3]$. The transformation thus also changes the
value of $r_\alpha $. One can check easily that
\begin{equation}
  \tilde r'_\alpha =\tilde r_\alpha +\ft13 (\partial ^if)k_{\alpha i},\qquad
  \omega '^i=\omega ^i-\ft12\rmi \kappa^2\partial ^if(\phi),\qquad {\cal P}'_\alpha ={\cal P}_\alpha.
 \label{KaTransSymm}
\end{equation}
The quantity $\omega ^i$ is thus the gauge field of these K{\"a}hler
transformations. On the other hand, $A_\mu ^B$ is as well the gauge field
for the part of the gauge transformations that act as R-symmetry,
(\ref{U1togauge}), as for the pull-back of the K{\"a}hler transformations to
spacetime, as
\begin{equation}
  A_\mu^{\prime B}=A_\mu ^B+\kappa^2\partial _\mu \Im f.
 \label{KaTransfA}
\end{equation}

A similar situation holds for $N=2$. In that case, one may rotate the
$\SU(2)$ vector quantities depending on a triplet of functions
$\vec{l}(q)$, and $\vec{\omega}_X$ is the gauge field for these
rotations:
\begin{equation}
  \delta _l \vec{J}_X{}^Y= \vec{l}\times\vec{J}_X{}^Y, \qquad \delta
  _l\vec{P}_I=\vec{l}\times\vec{P}_I, \qquad%
  \delta _l \vec{\omega }_X=-\ft12\partial _X\vec{l}+\vec{l}\times\vec{\omega
  }_X.
 \label{dellomega}
\end{equation}
The field $\vec{V}_\mu$,  (\ref{auxVmu}), is, apart from gauge field of
the $\SU(2)$ part of the gauge transformations, (\ref{LambdaSU2}), also
the gauge field for the pull-back to spacetime of these
$l$-reparametrizations:
\begin{equation}
  \delta _l\vec{{\cal V}}_\mu=\partial _\mu q^X\delta _l\vec{\omega }_X +
   g\kappa ^2A_\mu
  ^I\delta _l\vec{P}_I= -\ft12\partial_\mu \vec{l}(q(x)) + \vec{l}\times\vec{{\cal
  V}}_\mu.
 \label{dellVmu}
\end{equation}
Note that when one does not gauge any isometry on the K{\"a}hler manifold
($N=1$) or on the quaternionic-K{\"a}hler manifold ($N=2$), then there is no
$\Lambda _{\U(1)}$ or $\Lambda _{\SU(2)}$, but still the $A_\mu ^B$ or
$\vec{\cal V}_\mu$ gauges the reparametrizations on the scalar manifolds.

%
%

\begin{acknowledgement}
  Though much of the material in this review is known since a long time,
  the point of view that is presented here was stimulated mostly by
  discussions in recent collaborations. I thank E. Bergshoeff, P.
  Bin{\'e}truy, A. Celi, A. Ceresole, G. Dall'Agata, G. Dvali, J.
  Gheerardyn, R. Kallosh, S. Vandoren and M. Zagermann for related discussions.

This work is supported in part by the European Community's Human
Potential Programme under contract HPRN-CT-2000-00131 Quantum Spacetime,
and in part by the Federal Office for Scientific, Technical and Cultural
Affairs through the "Interuniversity Attraction Poles Programme --
Belgian Science Policy" P5/27.
\end{acknowledgement}

\providecommand{\href}[2]{#2}\begingroup\raggedright\endgroup

\end{document}